%% file: paper.tex
\documentclass[12pt]{article}
\usepackage{epsfig}

\usepackage{amsfonts}   
\usepackage{amssymb}
\usepackage{amsmath}    
\usepackage{graphicx}
\usepackage{caption}
\usepackage{subcaption}

\input econfmacros-csqcd4.tex
\def\Title#1{\begin{center} {\Large {\bf #1} } \end{center}}

\begin{document}

\Title{Non-Spherical Models of Neutron Stars}

\bigskip\bigskip


\begin{raggedright}  

{\it Omair Zubairi$^{1,2}$ ~~William Spinella$^{1,2}$ ~~ Alexis Romero$^2$
  ~~Richard Mellinger$^2$ ~~ Fridolin Weber$^{2,3}$ ~~ Milva
  Orsaria$^{4,5}$ ~~ and Gustavo Contrera$^{4,5}$ \\
  
  \bigskip
  
  $^1$Computational Science Research Center, San Diego State
  University, 5500 Campanile Drive, San Diego, CA 92182, USA

    \bigskip
  
  $^2$Department of Physics, San Diego State University, 5500
  Campanile Drive, San Diego, CA 92182, USA

    \bigskip
    
    $^3$Center for Astrophysics and Space Sciences, University of
    California San Diego, 9500 Gilman Drive, La Jolla, CA 92093, USA

    \bigskip
    
    $^4$Gravitation, Astrophysics and Cosmology Group, Facultad
  de Ciencias Astron{\'o}micas y Geof{\'i}sicas, Universidad
  Nacional de La Plata UNLP, La Plata, Argentina

      \bigskip
  
  $^5$CONICET, Rivadavia 1917, 1033 Buenos Aires, Argentina
    }
  
\end{raggedright}

\section{\label{sec:intro}Introduction}

Neutron stars are compact stellar objects with masses between around 1
and 2 solar masses and radii of around 10 to
15~km~\cite{blaschke01:trento,page06:review,becker08:a}.  They have
magnetic fields up to around $10^{18}$~G
\cite{thompson95:a,thompson96:a,thompson04:a,mereghetti08:a}.
Standard models for neutron stars traditionally assume that these
objects are perfect spheres whose properties are described, in the
framework of general relativity theory, by the well-known
Tolman-Oppenheimer-Volkoff (TOV) equation~\cite{oppenheimer39:a,
  tolman39:a}. The TOV equation is a simple first-order differential
equation which can be solved with little numerical effort (see, for
instance, Refs.\ \cite{glenn,weber}).

The assumption of perfect spherical symmetry may not be correct.  
It is known that magnetic fields are present in all neutron stars.  
In particular, if the magnetic field is strong (up to around
$10^{18}$~Gauss in the core) such as for magnetars~
\cite{thompson95:a,thompson96:a,thompson04:a,mereghetti08:a}, and/or
the pressure of the matter in the cores of neutron stars is
non-isotropic, as predicted by some models of color superconducting
quark matter~\cite{ferr}, then deformation of neutron stars can
occur \cite{chand,ferrv,goose,katz,hask,payne}.  We also mention the
recent work conducted by \cite{proto} which shows that high magnetic
fields in proto-quark stars modify quark star masses. The authors of this
study conclude that using the TOV equation would be insufficient for
numerical calculations of the properties of proto-quark stars.

The main goal of our study is to derive a TOV-like stellar-structure
equation for deformed neutron stars whose mathematical form is similar
to the standard TOV equation for spherical neutron stars.  This
equation will enable the user to explore the properties of deformed
neutron stars from an equation that can be solved with rather little
numerical effort, complementing more sophisticated numerical studies
such as the one presented very recently in \cite{mallick14:a}.

In contrast to the TOV stars that are composed of spherically
symmetric mass shells, the stellar models considered in our paper are
made of deformed mass shells which are either of oblate or prolate
shape. Strategically, such a treatment is similar to the formalism
developed by Hartle and Thorne \cite{hartle68:a}, which is based on a
quadrupole approximation of the metric of a rotating compact star. The
oblate and prolate shapes are obtained by parametrizing the polar
$(z)$ direction of the metric in terms of the equatorial $(r)$
direction along with a the deformation parameter $\gamma$, described
as $z={\gamma}r$, where we have assumed the symmetry to be axial
symmetric.  This parameter is normalized to $\gamma=1$ for a perfect
sphere.  An object that is deformed in the equatorial direction
(oblate spheroid) is obtained for $\gamma <1$, while an object
deformed in the polar direction (prolate spheroid) corresponds to
$\gamma > 1$.  Using this parametrization will allow us to keep the
energy momentum tensor in spherical form, while maintaining
deformation structure.

The parametrized metric allows us to derive the stellar structure
equation of deformed neutron stars in analytic form.  As already
mentioned above, this equation constitutes a generalization of the
well-known Tolman-Oppenheimer-Volkoff equation
\cite{oppenheimer39:a,tolman39:a}, which describes the properties of
perfect spheres in general relativity theory. 

The paper is organized as follows. In Sect.\ \ref{sec:eqs},
we discuss the derivation of the stellar structure equation of
deformed neutron stars (mathematical details are presented
in the Appendix).  The nuclear equation of state used to solve
this equation is introduced in Sect.\ \ref{sec:eos}.  Our equation
of state is based on a relativistic nuclear lagrangian which
describes confined hadronic matter and a nonlocal
Nambu-Jona-Lasinio lagrangian used to model
quark deconfinement.  The results are presented in
Sect.\ \ref{sec:results}.  They are of generic nature and do not
depend on the particular choice for the nuclear equation of
state.  Conclusions are drawn in Sect.\ \ref{sec:conc}.

\section{\label{sec:eqs}Stellar Structure Equations}

The properties of perfectly spherical stars are determined by the TOV
equation, which is based on the Schwarzschild metric given by
\begin{equation}
 \label{eq:schw_metric}
 ds^{2} = - {\rm e}^{2\Phi(r)} dt^{2} + {\rm e}^{2\Lambda(r)} dr^{2}
 + r^{2} d\theta^{2} + r^{2} \sin^{2}(\theta) d\phi^{2} \, ,
\end{equation}
where $t$ is the time coordinate, $r, \theta, \phi$ are the spatial
coordinates, and $\Phi(r)$ and $\Lambda(r)$ denote metric functions
which are determined from Einstein's field equation of general
relativity theory.  The deformed stellar models studied in this paper
are based on a metric that is similar to the one of
Eq.\ (\ref{eq:schw_metric}).  However, instead of spherical mass
shells the deformed stellar models are constructed from mass shells
that are either of prolate or oblate shape.  The mathematical form of
the metric of such objects reads ($G=c=1$)
\begin{eqnarray}
 \label{eq:gmetric}
 ds^{2} = g_{\mu\nu} dx^{\mu}dx^{\nu} = -{\rm e}^{2\Phi(r)} \, dt^{2}
 + \left(1-\frac{2m(r)}{r}\right)^{-\gamma} \, dr^{2} 
 +r^{2} \, d\theta^{2} + r^{2} \sin^{2}(\theta) \, d\phi^{2} \, ,
\end{eqnarray}
where $\gamma$ denotes a constant that determines the degree of
deformation.  To derive the hydrostatic equilibrium equation
associated with Eq.\ (\ref{eq:gmetric}), we start with Einstein's field
equation in the mixed tensor representation
\begin{equation}
\label{eq:ein}
 G^\mu{}_\nu {\equiv} R^\mu{}_\nu - \frac{1}{2} \, R \, g^\mu{}_\nu =
 - 8 {\pi} T^\mu{}_\nu \, .
\end{equation}
Here $G^\mu{}_\nu$ denotes the Einstein tensor, which is given in
terms of the Ricci tensor $R^\mu{}_\nu$, the Ricci scalar $R$ and the
metric tensor $g^\mu{}_\nu= \delta^\mu{}_\nu$.  The energy-momentum
tensor
\begin{equation}
 \label{eq:enmom}
T^\mu{}_\nu=(\epsilon + P) \, u^\mu \, u_\nu  - g^\mu{}_\nu \, P 
\end{equation}
is given in terms of the stellar equation of state (pressure, $P$, as
a function of energy density, $\epsilon$) and the matter's
four-velocity $u^\mu= dx^\mu / d\tau$ and $u_\nu= dx_\nu / d\tau$,
with the proper time $\tau$ given by $d\tau^2 = ds^2$.  Using
Eqs.~(\ref{eq:gmetric}) through (\ref{eq:enmom}) along with the
equations provided in the Appendix, one arrives at
the stellar structure equation of a deformed neutron star,
\begin{equation}
\label{eq:gtov}
 \frac{dP}{dr}=-\frac{(\epsilon+P)\left[\frac{1}{2}r+4{\pi}r^{3}P
     -\frac{1}{2}r \left(1-\frac{2m}{r}\right)^{\gamma}\right]}
      {r^{2}\left(1-\frac{2m}{r}\right)^{\gamma}} \, .
\end{equation}
In the limiting case when $\gamma=1$, Eq.\ (\ref{eq:gtov}) becomes the
well-known Tolman-Oppenheimer-Volkoff
equation~\cite{oppenheimer39:a,tolman39:a}
\begin{equation}
 \label{eq:ogtov}
 \frac{dP}{dr}=-\frac{\left(\epsilon+P\right)
   \left(m+4{\pi}Pr^{3}\right)} {r^{2}\left(1 - \frac{2m}{r}\right)}
 \, ,
\end{equation}
which describes the structure of perfectly spherically symmetric
objects. The gravitational mass of a deformed neutron star is given by
\begin{equation}
 \label{eq:gmass}
 \frac{dm}{dr}=4 {\pi} r^2 {\epsilon} {\gamma} \, ,
\end{equation}
so that the total gravitational mass, $M$, of a deformed neutron star
with an equatorial radius $R$ follows as~\cite{herr}
\begin{equation}
\label{eq:ggm}
 M=\gamma\, {m(R)} \, .
\end{equation}
In the spherical limit, the total mass is given by $M
\equiv m(R) = 4 \pi \gamma \int_0^R dr r^2 \epsilon$.  The stellar
radius $R$ is defined by the condition that pressure at the surface of
a neutron stars vanishes, that is, $P(r=R)=0$.

It is important to investigate the space outside the star as well.
For that we need to examine the ${\rm e}^{2\Phi(r)}$ component of
Eq.~(\ref{eq:gmetric}).  Using the equations given in the Appendix,
one finds
\begin{equation}
\label{eq:dphidr}
\frac{d{\Phi}}{dr}=\frac{\left[\frac{1}{2}+4{\pi}r^{2}P -\frac{1}{2}
    \left(1-\frac{2m}{r}\right)^{\gamma}\right]}{r\left(1 -
  \frac{2m}{r}\right)^{\gamma}} \, .
\end{equation}
One can easily see from Eq.~(\ref{eq:dphidr}) that asymptotically
$d{\Phi}/dr \rightarrow 0$, as required.

Now that we are equipped with the stellar structure equations
(\ref{eq:gtov}) and (\ref{eq:ggm}), which are dependent on the
deformation parameter $\gamma$, we solve them for a given equation of
state.  The model chosen here assumes that neutron stars are made of
quark-hybrid matter. It is based on a relativistic nuclear lagrangian
to describe confined hadronic matter and a nonlocal Nambu-Jona-Lasinio
lagrangian to model quark matter~\cite{njl2014,njlcp}.  Phase
equilibrium in the quark-hadron mixed phase is governed by the Gibbs
condition.  Section~\ref{sec:eos} briefly describes the key features
of this equation of state. We stress, however, that the results
presented in Sect.~\ref{sec:results} are generic and do not depend on
the particular choice for the nuclear equation of state.

\section{\label{sec:eos}Equation of State}

\subsection{\label{subsec:eos1}Hadronic Matter}

At densities higher than that of the inner neutron star crust, and
lower than required for quark deconfinement, we model neutron star
matter composed of baryons
($B=\{n,p,\Lambda,\Sigma,\Xi,\Delta,\Omega\}$) and leptons
($\lambda=\{e^-,\mu^-\}$) using the relativistic mean-field
approximation.  The Lagrangian is given by
\cite{glenn,weber,selfinteractions}
\begin{eqnarray}
\label{eq:hlag}
  \mathcal{L} &=& \sum\limits_{B}
    \bar\psi_B\big[\gamma_{\mu}(i\partial^{\mu}
    -g_{\omega}\omega^{\mu}-g_{\rho}\vec{\boldsymbol{\tau}}\cdot
    \vec{\boldsymbol{\rho}}^{\mu})
    -(m_N-g_{\sigma}\sigma)\big]\psi_B
    +\frac{1}{2}(\partial_{\mu}\sigma\partial^{\mu}\sigma-
    m^2_{\sigma}\sigma^2)\nonumber\\
    &&-\frac{1}{3}b_{\sigma}m_N
    (g_{\sigma}\sigma)^3-\frac{1}{4}c_{\sigma}
    (g_{\sigma}\sigma)^4-\frac{1}{4}\omega_{\mu\nu}
    \omega^{\mu\nu}
    +\frac{1}{2}m^2_{\omega}\omega_{\mu}\omega^{\mu}+\frac{1}{2}
    m^2_{\rho}\vec{\boldsymbol{\rho}}_{\mu}\cdot
    \vec{\boldsymbol{\rho}}^{\mu} \nonumber\\
    &&-\frac{1}{4}\vec{\boldsymbol{\rho}}_{\mu\nu}\cdot
    \vec{\boldsymbol{\rho}}^{\mu\nu}
    +\sum\limits_{\lambda}\bar\psi_{\lambda}(i\gamma_{\mu}
    \partial^{\mu}-m_\lambda)\psi_{\lambda} \, .
\end{eqnarray}
The interactions between baryons are described by the exchange of
scalar, vector, and isovector mesons
\begin{table}[tbh]
  \begin{center}
    \begin{tabular}{lc}
      Saturation Properties~~~&~~~ NL3 Parametrization \\
      \hline
      $\rho_0$ (fm$^{-3}$) & 0.148 \\
      $E/N$ (MeV) & $-16.3$ \\
      $K$ (MeV) & 272 \\
      $m^*/m_N$ & 0.60 \\
      $a_{sy}$ (MeV) & 37.4 \\
      \hline
    \end{tabular}
    \caption{Parametrization of hadronic matter, where the saturation
  properties are baryonic density $\rho_{0}$, energy per baryon $E/N$,
  nuclear incompressibility $K$, effective nucleon mass $m^{*}_{N}$,
  and asymmetry energy $a_{\rm sy}$.}
    \label{table:nl3}
  \end{center}
\end{table}
($\sigma,\omega,\rho$)~\cite{walecka}.  In the present work we employ
the NL3 parametrization as given in Table \ref{table:nl3}
\cite{nl3}. For further details, see Refs.\ \cite{glenn,weber} and
references therein.

\subsection{\label{subsec:eos2}Quark Matter}

To determine the equation of state of the deconfined quark phase we
use a nonlocal extension of the three-flavor Nambu-Jona-Lasinio model
(see \cite{njl2014} and references therein). This model hosts numerous
improvements over other models of deconfined quark matter, including
but not limited to the treatment of vector interactions among quarks,
reproduction of confinement for proper parametrization, lack of
ultraviolet divergences with the introduction of the nonlocal form
factor $g(\tilde z)$, and momentum dependent dynamical quark masses.
The Euclidean effective action is given by
\begin{eqnarray}
  \label{eq:action}
    S_E &=&\int d^4x\, \Big\{\bar\psi(x)\big[-i\gamma_{\mu}
          \partial^{\mu}+\hat{m}\big]\psi(x)
          -\frac{G_S}{2}\big[j_a^S(x)j_a^S(x)-j_a^P(x)j_a^P(x)\big]  \\
          &&-\frac{H}{4}T_{abc}\big[j_a^S(x)j_b^S(x)j_c^S(x)-
          3j_a^P(x)j_b^P(x)j_c^P(x)\big]
          -\frac{G_V}{2}j_{V,f}^{\mu}(x)j_{V,f}^{\mu}(x)\Big\} \, ,
          \nonumber 
\end{eqnarray}
where $\psi=(uds)^T$, $\hat{m}=\mathrm{diag}(m_u,m_d,m_s)$, and $H$,
$G_S$, and $G_V$ are coupling constants.  For convenience we assume
$m_u=m_d=\bar{m}$.  The scalar, pseudo-scalar, and vector currents are
respectively
\begin{equation}
  j_{a}^{S}(x) = \int d^4z\,\tilde{g}(z)\bar\psi\left(x+\frac{z}{2}\right)
  \lambda_a\psi\left(x-\frac{z}{2}\right) \, ,
\end{equation}
\begin{equation}
  j_{a}^{P}(x) = \int d^4z\,\tilde{g}(z)\bar\psi
  \left(x+\frac{z}{2}\right)
  i\gamma_5\lambda_a\psi\left(x-\frac{z}{2}\right) \, ,
\end{equation}

\begin{equation}
  j_{V}^{\mu}(x) = \int d^4z\,\tilde{g}(z)\bar\psi
  \left(x+\frac{z}{2}\right)
  \gamma^{\mu}\lambda_a\psi\left(x-\frac{z}{2}\right) \, .
\end{equation}
Applying standard bosonization to (\ref{eq:action}) we derive the
thermodynamic potential in the mean-field approximation at zero
temperature \cite{njl2014}.  We use the same parametrization for
the nonlocal NJL model as given in Ref. \cite{njl2014}. The vector
coupling constant ($G_V$) is given in terms of the scalar coupling
constant ($G_S$) and is chosen to be $G_V = 0.09~G_S$.

\subsection{\label{subsec:eos3}Quark-Hadron Mixed Phase} 

Phase equilibrium between the hadronic and quark phases of
neutron star matter is governed by the Gibbs condition,
\begin{equation}
  p_H(\mu_n,\mu_e,\{\phi\}) = p_Q(\mu_n,\mu_e,\{\psi\}),
\end{equation}
where $\mu_n$ and $\mu_e$ are the neutron and electron chemical
potentials, and $\{\phi\}$ and $\{\psi\}$ are the field variables and
Fermi momenta associated with solutions of the equations of hadronic
and quark matter, respectively.  When this condition is initially met
the first order phase transition from hadronic to quark matter
begins. The relaxed condition of global charge neutrality allows the
hadronic matter to become more isospin symmetric by transferring
negative charge from the hadronic to the quark phase, lowering the
asymmetry energy. This results in a mixed phase with coexisting
regions of positively charged hadronic matter and negatively charged
quark matter \cite{glenn,ng92,ngpr}. The equation of state for this
phase is solved by combining the approaches for hadronic and quark
matter under the Gibbs condition, baryon number conservation, and
global electric charge neutrality.

\section{\label{sec:results}Results}

\begin{figure}[ht]
\begin{center}
\epsfig{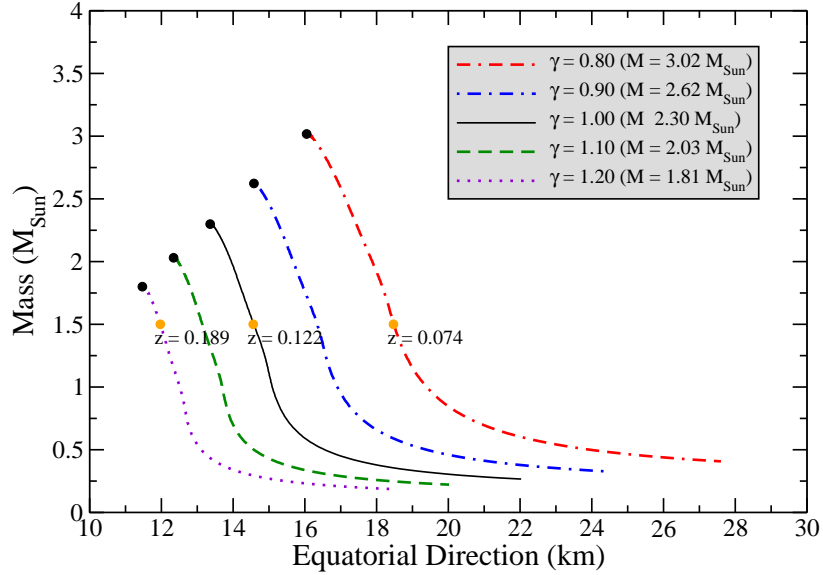}
\caption{(Color online) Mass-radius relationships of deformed neutron
  stars ($\gamma<1$: oblate neutron stars, $\gamma >1$: prolate
  neutron stars, $\gamma=1$: spherical neutron stars).  The solid dots
  on each curve represent the maximum-mass star for each stellar
  sequence.}
\label{fig:njlmvsr}
\end{center}
\end{figure}

\begin{figure*}[htb]
    \centering
    \begin{subfigure}[t]{0.33\textwidth}
        \centering
       \includegraphics[width=1.00\textwidth]{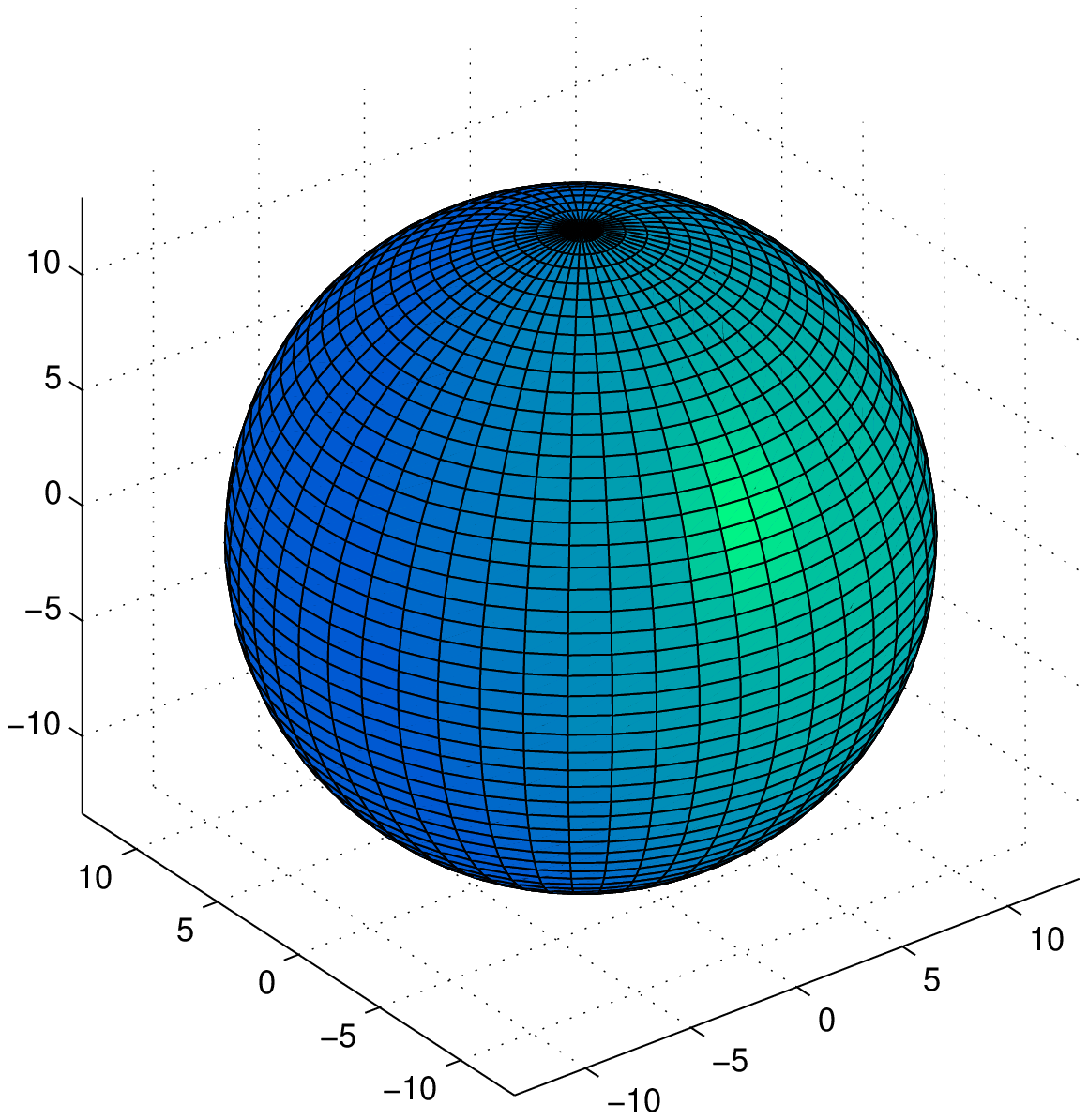}
        \caption{$\gamma=1.00$; ${\rm M}=2.30~{\rm M}_{\odot}$}
    \end{subfigure}%
    \begin{subfigure}[t]{0.33\textwidth}
        \centering
        \includegraphics[width=1.00\textwidth]{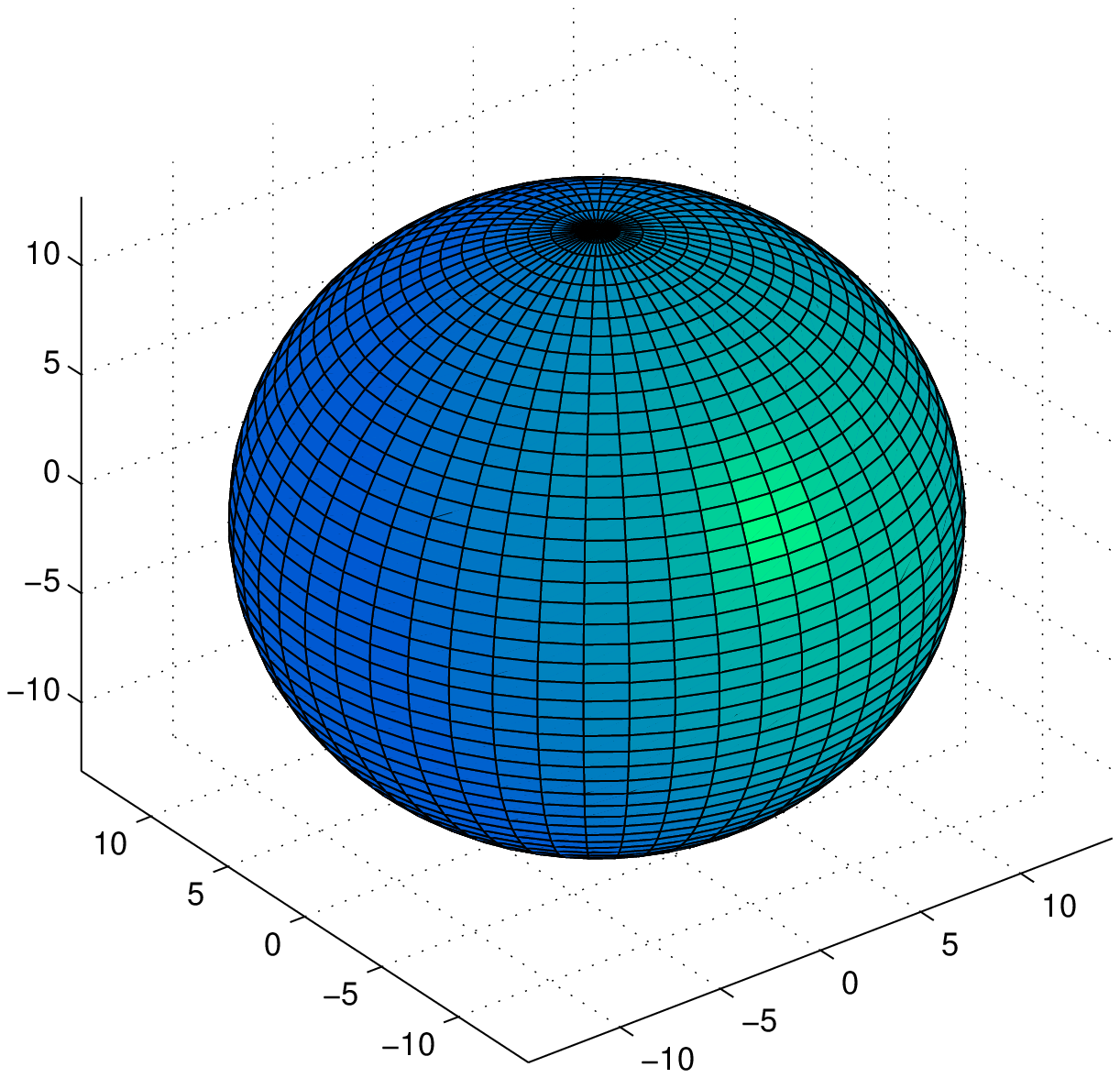}
        \caption{$\gamma=0.90$; ${\rm M}=2.62~{\rm M}_{\odot}$}
    \end{subfigure}
    \begin{subfigure}[t]{0.33\textwidth}
        \centering
        \includegraphics[width=1.00\textwidth]{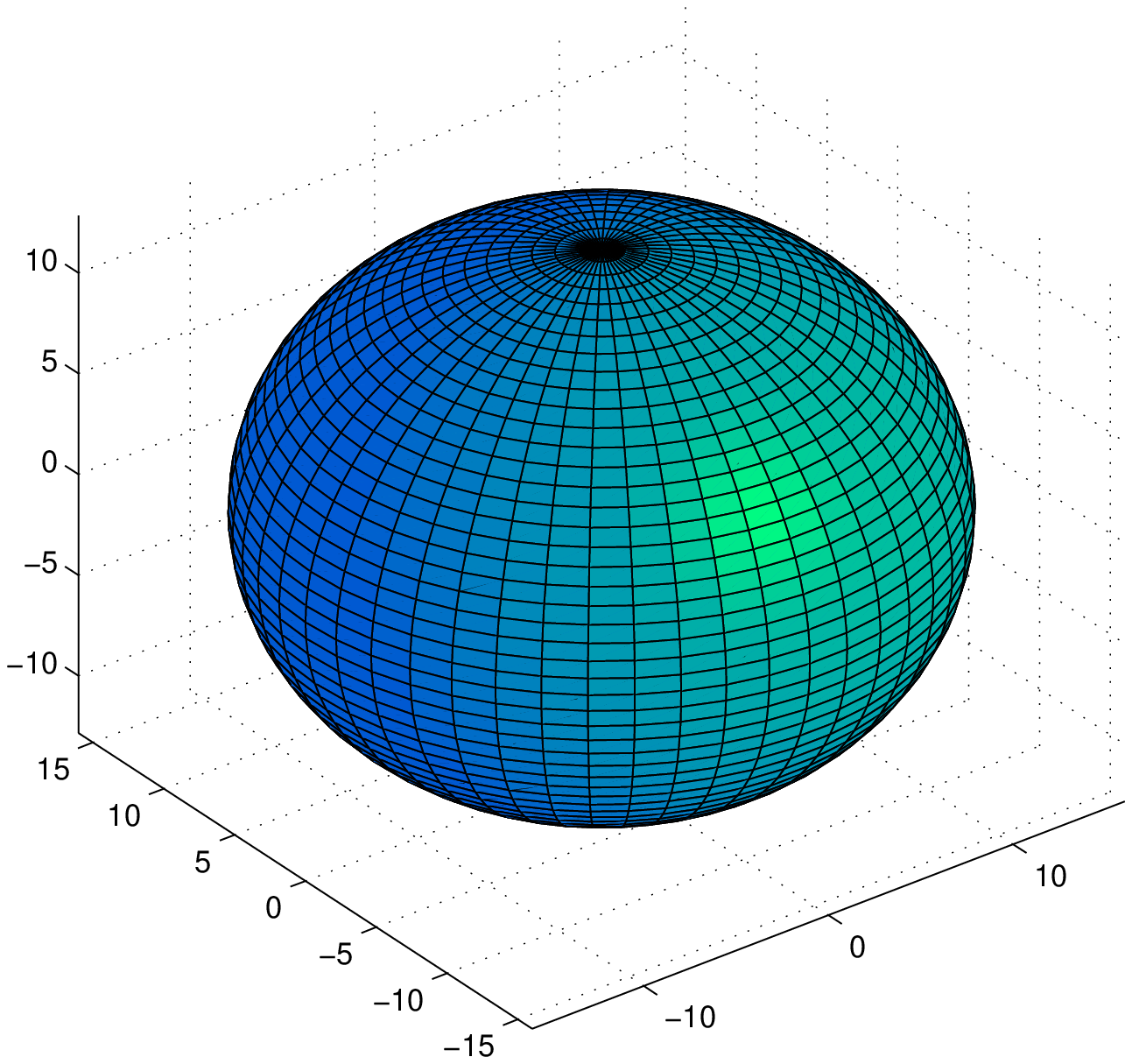}
        \caption{$\gamma=0.80$; ${\rm M}=3.02~{\rm M}_{\odot}$}
    \end{subfigure}    
    \caption{Shapes of the oblate maximum-mass stars shown in Fig.~\ref{fig:njlmvsr}.}
    \label{fig:obmass}
\end{figure*}

\begin{figure*}[htb]
    \centering
    \begin{subfigure}[t]{0.33\textwidth}
        \centering
        \includegraphics[width=1.00\textwidth]{gama1-00.eps}
        \caption{$\gamma=1.00$; ${\rm M}=2.30~{\rm M}_{\odot}$}
    \end{subfigure}%
    \begin{subfigure}[t]{0.33\textwidth}
        \centering
        \includegraphics[width=1.00\textwidth]{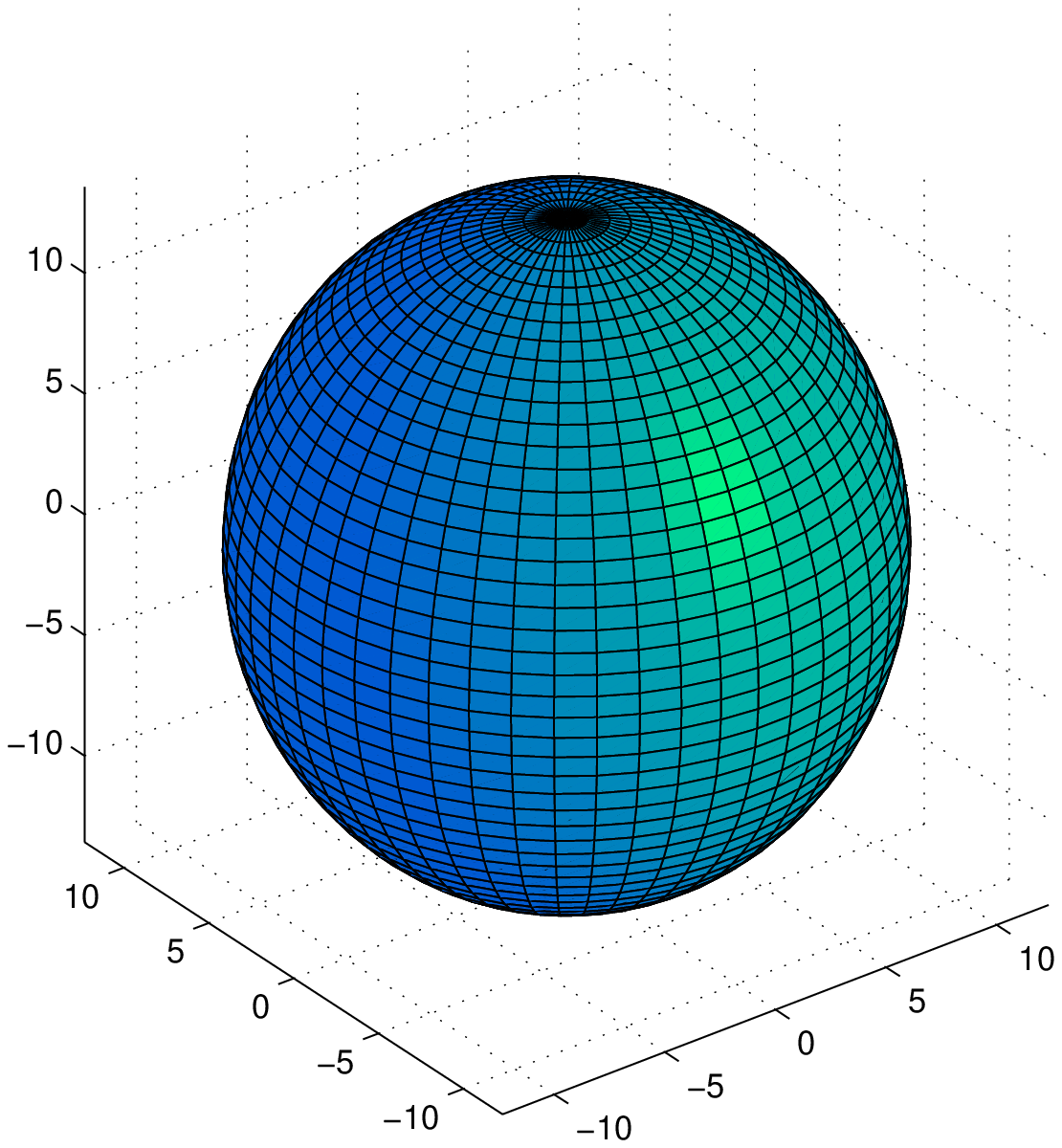}
        \caption{$\gamma=1.10$; ${\rm M}=2.03~{\rm M}_{\odot}$}
    \end{subfigure}
    \begin{subfigure}[t]{0.33\textwidth}
        \centering
        \includegraphics[width=1.00\textwidth]{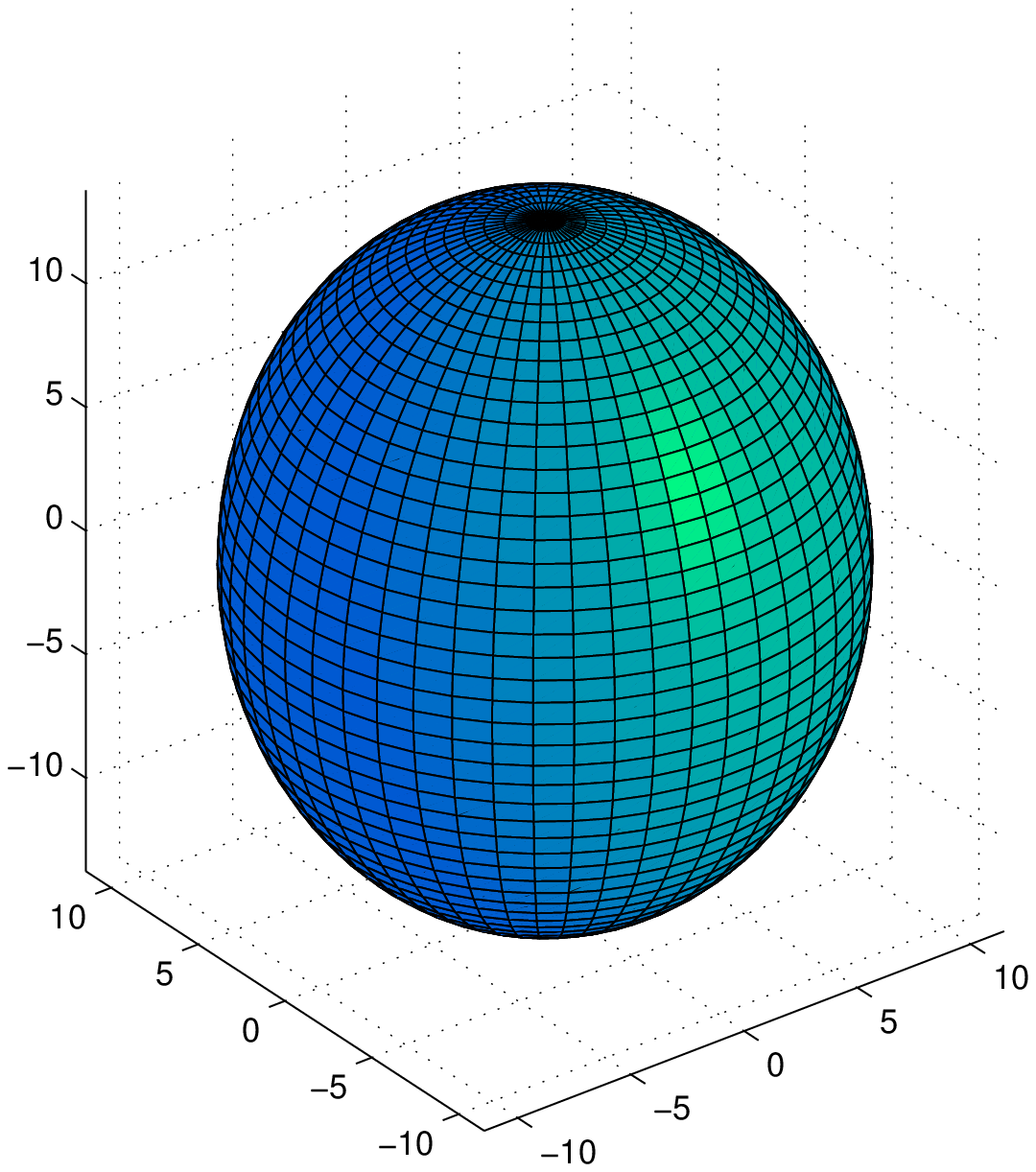}
        \caption{$\gamma=1.20$; ${\rm M}=1.81~{\rm M}_{\odot}$}
    \end{subfigure}    
    \caption{Shapes of the prolate maximum-mass stars shown in
      Fig.~\ref{fig:njlmvsr}.}
    \label{fig:promass}
\end{figure*}

We first calculate the masses and radii of non-spherical neutron stars
by solving Eqs.~(\ref{eq:gtov}) and (\ref{eq:gmass}) numerically using
the Runge Kutta method.  The outcome is shown in
Fig.~\ref{fig:njlmvsr}.
From the results shown in this figure, we see that the maximum mass of
the spherical ($\gamma=1$) neutron star obtained for the equation of
state of this work is $2.3~{\rm M}_{\odot}$. The equatorial radius of
this star is close to 14 kilometers~\cite{njl2014}.  Oblate neutron
stars are obtained for $\gamma$ values less than one, since the polar
coordinate obeys $z=\gamma r$. We find that a decrease of $\gamma$ by
10\% results in a $\sim15\% $ increase in gravitational mass and an
increase in equatorial radius by a few kilometers
(Fig.\ \ref{fig:njlmvsr}).  If we continue decreasing $\gamma$, the
mass keeps increasing monotonically, ultimately extending into the
mass region of solar-mass black holes. Prolate neutron stars are
obtained for deformation parameters $\gamma~>1$.  In this case, as
shown in Fig.~\ref{fig:njlmvsr}, an increase of $\gamma$ by 10\% leads
to a $\sim 12\%$ decrease in gravitational mass and a decrease in
equatorial radius. If one keeps decreasing $\gamma$ further, the
\begin{figure}
\centering
\includegraphics[width=0.70\textwidth]{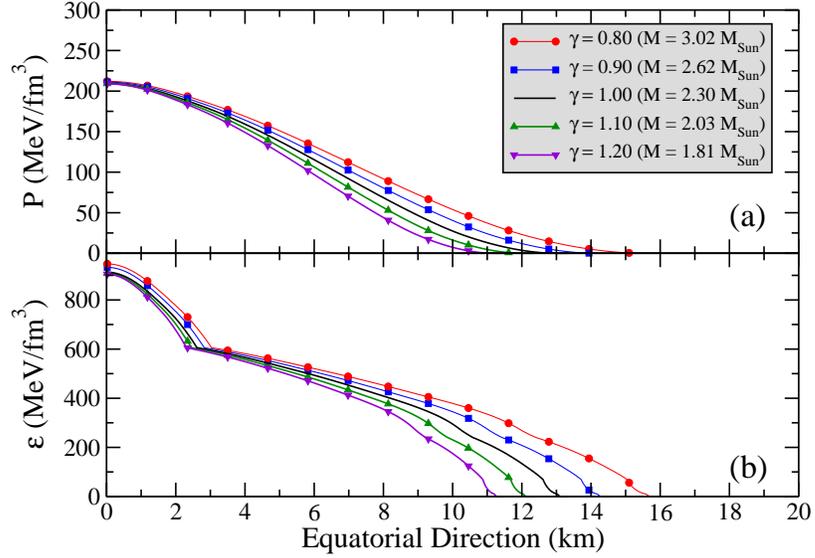}
\caption{(Color online) (a) Pressure profiles and (b) energy-density
  profiles in equatorial direction for the maximum-mass stars shown
  Fig.~\ref{fig:njlmvsr}.}
\label{fig:pdvsr}
\end{figure}
maximum mass of a deformed neutron star drops down toward the $1.5~
{\rm M}_\odot$ region.  The situation is graphically illustrated in
Figs.~\ref{fig:obmass} and \ref{fig:promass}, which show the
deformations of the maximum-mass neutron stars of
Fig.~\ref{fig:njlmvsr} for $\gamma$ values ranging from 0.8 to 1.0
\begin{figure}
\bigskip
 \centering
\includegraphics[width=0.70\textwidth]{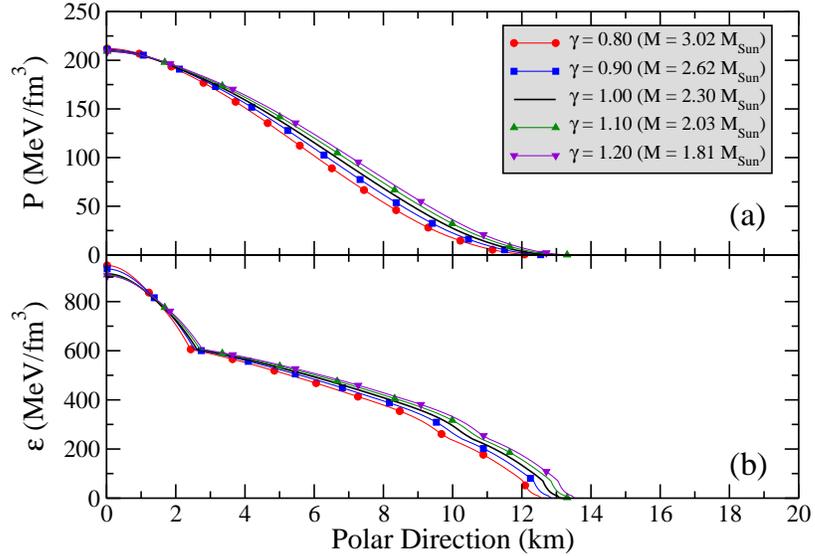}
\caption{(Color online) Same as Fig.~\ref{fig:pdvsr} but in polar
  direction.}
\label{fig:pdvsz}
\end{figure}
(oblate to spherical deformations) to 1.0 to 1.2 (spherical to prolate
deformations).  The pressure and energy-density profiles of the
maximum-mass neutron stars of Fig.~\ref{fig:njlmvsr} are shown in
Figs.~\ref{fig:pdvsr} and \ref{fig:pdvsz}.

The bottom line of all this is that there may be multiple maximum-mass
neutron stars for one and for the same model for the nuclear equation of
state, depending on the type (oblate or prolate) of stellar deformation,
which in the end is linked to the strengths of the magnetic fields of neutron
stars and/or anisotropic pressure gradients in their cores. Moreover, as
indicated by our calculations, the deformation does not need to be very large to
appreciably change the bulk properties of neutron stars.  This finding
may be critical to better understand the ever-widening range of
observed neutron star masses and to discriminate neutron stars from
solar-mass black holes.

Next, we calculate the energy loss of photons emitted from the surface
of a deformed neutron star.  We consider a photon created at the
surface of the star (emitter) and leaving its gravitational field
toward a detector located at infinity, where space-time is flat. The
photon's frequency at the emitter, $\nu_E$, is given as the inverse of
the proper time between two wave crests, $d \tau_E$, that is, $\nu_E =
1/ d \tau_E = (- g_{\mu\nu} d x^\mu d x^\nu)_E^{-1/2}$, where $d x^1 =
d x^2 = d x^3=0$ because the emitter stays at a fixed position while
emitting the photon. The same expression written down for the receiver
at infinity reads $\nu_\infty = 1 / d \tau_\infty = (- g_{\mu\nu} d
x^\mu d x^\nu)_\infty^{-1/2}$. The ratio of these two frequencies is
given by
\begin{eqnarray}
  \frac{\nu_\infty}{\nu_E} = \frac{[(- g_{00})^{1/2} d x^0]_E}{[(-
    g_{00})^{1/2} d x^0]_\infty}  \, .
\label{eq:freq.infinity}  
\end{eqnarray}
If we assume that the coordinate time $d x^0$ between two wave
crests is the same as at the star's surface and the receiver, which is
the case if the gravitational field is static so that whatever the
world-line of one photon is from the star to the receiver, the next
photon follows a congruent path, merely displaced by $d x^0$ at all
points \cite{glenn,weber}, this ratio simplifies to $\nu_\infty/\nu_E
= [(- g_{00})^{1/2}]_E / [(- g_{00})^{1/2}]_\infty$.  Making use of
the definition of the gravitational redshift, $z = (\nu_E/\nu_\infty)
- 1$, we obtain
\begin{equation}
  z = \Bigl( 1 - \frac{2\, M}{R} \Bigr)^{-\gamma/2} - 1 \, .
\label{eq:redshift}
\end{equation} 
Equation (\ref{eq:redshift}) shows that the gravitational redshift
carries important information about the mass, radius, and the
deformation of a neutron star. The $z$ values at the equators of
several deformed $1.5~ {\rm M}_\odot$ neutron stars are shown in
Fig.~\ref{fig:njlmvsr}.

The eccentricities $e$,
\begin{eqnarray}
  e \equiv {\rm sign} (R_{\rm eq} - R_{\rm p})\,  \sqrt{ 1 - \left(
  \frac{R^{<}}{R^{>}} \right)^2 }~, 
\label{eq:eccentricity}
\end{eqnarray}
of the neutron stars shown in Figs.~\ref{fig:obmass} and
\ref{fig:promass} are summarized in Table \ref{table:e}.  For
spherical neutron stars
$R^<$ (semi-minor axis) and $R^>$ (semi-major~axis) are
equal, so that $e=0$ for such objects. Neutron stars whose $\gamma$
values differ by $\pm 10$\% from the spherical case have
eccentricities of $e=0.43617$ if the deformation is oblate and
$e=-0.41601$ if the deformation is prolate. Rapid rotation also
deforms neutron stars away from spherical symmetry. For the neutron
stars of this paper, we find eccentricities as low as 0.6 for rotation
at the mass shedding frequency (which sets an absolute limit on rapid
rotation), but not smaller. The metric of a rotating neutron star can
also be used to study the structure of deformed non-rotating neutron
stars. This has been done recently in Ref.\ \cite{mallick14:a}.  The
results of this paper cannot be directly compared with our results,
however, because of specific assumptions about the energy-momentum
tensor. We note, however, that the eccentricities of the oblate
neutron stars of our study are compatible with those obtained in
\cite{mallick14:a}, depending on the degree of anisotropy generated by
the magnetic field.

\section{\label{sec:conc}Conclusions}
  
The goal of this work was to investigate the impact of deformation on
the structure of non-rotating neutron stars in the framework of
general relativity.  For this purpose we first derived a stellar
structure equation that describes deformed neutron stars. This
equation constitutes a generalization of the well-known
Tolman-Oppenheimer-Volkoff (TOV) equation, which describes the
structure of non-rotating, perfectly spherically symmetric neutron
stars.  The mathematical structure of this generalized TOV equation is
such that the deformation of a neutron star (or any other compact
object, such as a hypothetical quark star) is expressed in terms of a
deformation parameter, $\gamma$.  By virtue of this parameter, models
of deformed neutron stars can be built from non-spherical (prolate or
oblate) mass shells rather than spherical mass shells. This leads to a
stellar structure equation for deformed neutron stars which is of the
same simple mathematical structure as the standard TOV equation and
thus can be solved with little numerical effort.

The parametrization introduced in our paper allows one to use a model
for the equation of state in the limiting case of isotropy while
maintaining deformation structure.  From the results shown in
Fig.~\ref{fig:njlmvsr}, one sees that modest
\begin{table}[htb]
 \begin{center}
 \begin{tabular}{cccc} 
   &$\gamma=0.80$  ~~~~~~~&$\gamma=0.90$~~~~~~~  &$\gamma=1.00$ \\ \cline{2-4}
   $\epsilon$ &$0.60000$ &$0.43617$ &$0$ \\ \hline
   &$\gamma=1.00$  ~~~~~~~&$\gamma=1.10$~~~~~~~  &$\gamma=1.20$ \\ \cline{2-4}
   $\epsilon$ &$0$ &$-0.41601$ &$-0.55222$ \\
   \hline
 \end{tabular}
 \caption{Eccentricities, $\epsilon$, of the oblate and prolate neutron
  stars shown in Figs.~\ref{fig:obmass} and \ref{fig:promass},
  respectively.} \label{table:e}   
\end{center}
\end{table}
deformations can lead to appreciable changes in a neutron star's
gravitational mass and radius. In particular, we find that the mass of
a neutron star increases with increasing oblateness, but decreases
with increasing prolateness. This opens up the possibility that,
depending on the degree of stellar deformation, there may exist multiple
maximum-mass neutron stars for one and the same model for the nuclear
equation of state, which is drastically different for spherically
symmetric neutron stars whose mass-radius relationships are
characterized by one and only one maximum-mass star. This finding may
be critical to properly understand the ever widening range of observed
neutron star masses and to discriminate neutron stars from solar-mass
black holes.

\section*{Acknowledgments}

This work is supported through the National Science Foundation under
grants PHY-1411708 and DUE-1259951.  A. Romero is supported by NIH
through the Maximizing Access to Research Careers (MARC), grant number
5T34GM008303-25.  Computing resources are provided by the
Computational Science Research Center and the Department of Physics at
San Diego State University.
The authors would like to thank Vivian de la Incera and Efrain Ferrer (UTEP)
for their insightful discussions and initial motivation on this work.

\section{Appendix}{\label{sec:mathapp}}

Below, we outline the derivation of the stellar structure equation of
deformed compact objects.  For the metric given in
Eq.~(\ref{eq:gmetric}), the non-vanishing Christoffel symbols are
\begin{eqnarray}
 \label{eq:chriss}
\Gamma^{r}_{~tt} ={\beta}~{\rm e}^{2\Phi(r)}\Phi^{\prime}(r) \, , ~~
\Gamma^{t}_{~tr} =\Phi^{\prime}(r) \, ,~~
\Gamma^{r}_{~rr}=\dfrac{{\gamma}\left[-m^{\prime}
    (r)r+m(r)\right]}{r[-r+2m(r)]}\, , ~~ \Gamma^{r}_{~\theta\theta} =-{\beta}~r \, , ~~
\nonumber \\
\Gamma^{\theta}_{~r\theta} =\Gamma^{\Phi}_{~r\Phi}=\dfrac{1}{r} \,  , ~~
\Gamma^{\Phi}_{~\Phi\Phi} =\cot(\theta) \, , ~~
\Gamma^{r}_{~\Phi\Phi} =-{\beta}~r\sin^{2}(\theta) \, , 
\Gamma^{\theta}_{~\theta\theta} = -\sin(\theta)\cos(\theta) \, , ~~
\end{eqnarray}
where primes denote derivatives with respect to the radial coordinate, $r$, and
\begin{equation}
\beta{\equiv}\left(\frac{r-2m(r)}{r}\right)^{\gamma} \, .
\end{equation}

\noindent
The components of the Ricci tensor $R^{\mu}{}_{\nu}$ for the metric of
Eq.~(\ref{eq:gmetric}) are calculated to be
\begin{eqnarray}
\label{eq:riccitt}
R^{t}{}_{t}&=&
\frac{1}{r(r-2m(r))}\Big[\beta\Phi^{\prime}(r)m^{\prime}(r)\gamma{r} -
  \Phi^{\prime}(r)m(r)\gamma -\left(\Phi^{\prime}(r)\right)^{2} r\, m(r)
  -\Phi^{\prime\prime}(r) r^{2}
\nonumber\\
 &&
  + 2\Phi^{\prime\prime}(r)r \, m(r)
  -2\Phi^{\prime}(r)r+4\Phi^{\prime}(r)m(r)\Big] \, ,
\end{eqnarray}

\begin{eqnarray}
\label{eq:riccirr}
R^{r}{}_{r}&=&\frac{1}{r^{2}(r-2m(r))}\Big[-\beta\Phi^{\prime\prime}(r)r^{3}
  - 2\Phi^{\prime\prime}(r)r^{2}m(r)
  +\left(\Phi^{\prime}(r)\right)^{2}r^{3} -
  2\left(\Phi^{\prime}(r)\right)^{2}r^{2}m(r)\nonumber
  \\ &&-\gamma\Phi^{\prime}(r)m^{\prime}(r)r^{2}
  +\gamma\Phi^{\prime}(r)r\, m(r)-2{\gamma}m^{\prime}(r)r+2{\gamma}m(r)\Big] \, ,
\end{eqnarray}
\begin{eqnarray}
\label{eq:riccitheta}
R^{\theta}{}_{\theta} &=& \frac{1}{r^{2}(r-2m(r))}\Big[-{\beta}r^{2}\Phi^{\prime}(r)
  +2\beta\gamma{r}\Phi^{\prime}(r)m(r)
  +\beta{\gamma}m^{\prime}(r)r-\beta{\gamma}m(r)\nonumber
  \\ &&+\, r-2m(r)-\beta\gamma+2{\beta}m(r)\Big] \, ,
\end{eqnarray}
and 
\begin{equation}
\label{eq:ricciphi}
R^{\phi}_{\phi}=R^{\theta}_{\theta} \, .
\end{equation}
The Ricci scalar, $R$, is calculated to be
\begin{eqnarray}
\label{eq:ricciscalar}
R &=&\frac{2}{r^{2}(r-2m(r))} \left[\beta\gamma\Phi^{\prime}(r)m^{\prime}(r)r^{2}
  - \beta\gamma\Phi^{\prime}(r)m(r)r
  -\beta\left(\Phi^{\prime}(r)\right)^{2}r^{3}\right. \nonumber  \\
  &&+ 2\beta\left(\Phi^{\prime}(r)\right)^{2}r^{2}m(r)-\beta\Phi^{\prime\prime}(r)r^{3}
   +2\beta\Phi^{\prime\prime}(r)r^{2}m(r)-2m(r)
  -2\beta\gamma^{2}\Phi^{\prime}(r) \nonumber \\ && +
  4\beta\Phi^{\prime}(r)m(r)+2\beta{\gamma}m^{\prime}(r)r
  -2\beta{\gamma}m(r) +r -2m(r)-\beta\gamma+2{\beta}m(r)\Big] \, . ~~~~~
\end{eqnarray}
Substituting Eqs.~(\ref{eq:riccitt}) to~(\ref{eq:ricciphi}) along with
Eq.~(\ref{eq:ricciscalar}) into Einstein's field equation (\ref{eq:ein}),
one arrives at the general relativistic stellar structure equation
(\ref{eq:gtov}) of deformed compact objects.

\end{document}

%% file: econfmacros-csqcd4.tex
\def\beq{\begin{equation}}
\def\eeq#1{\label{#1}\end{equation}}
\def\eeqn{\end{equation}}

\def\beqa{\begin{eqnarray}}
\def\eeqa#1{\label{#1}\end{eqnarray}}
\def\eeqan{\end{eqnarray}}

\let\bar=\overbar

\def\Dslash{\not{\hbox{\kern-4pt $D$}}}
\def\dslash{\not{\hbox{\kern-2pt $\del$}}}

\def\msb{{\bar{\ssstyle M \kern -1pt S}}}

\usepackage{fancyhdr,graphicx}